\documentclass[12pt]{iopart}
\usepackage{epsfig}
\usepackage{graphicx}
\usepackage{amssymb}
\usepackage{color}
\begin{document}
\DeclareGraphicsExtensions{.eps, .jpg}
\input epsf

\title {An extended infrared study of the {$p,T$} phase diagram of the $p$-doped Cu-O plane}
\author{D. Nicoletti$^{1}$, P. Di Pietro$^{1}$,  O. Limaj$^{2}$, P. Calvani$^{1}$, U. Schade$^{3}$ S. Ono$^{4}$, Yoichi Ando$^{5}$, and S. Lupi$^{2}$}
\address{$^{1}$CNR-SPIN and Dipartimento di Fisica,  Universit\`{a} di Roma La Sapienza, Piazzale A. Moro 2, I-00185 Roma, Italy}
\address{$^{2}$ CNR-IOM and Dipartimento di Fisica, Universit\`a di Roma La Sapienza, Piazzale A. Moro 2, I-00185, Roma, Italy}
\address{$^{3}$Berliner Elektronenspeicherring-Gesellshaft f\"ur Synchrotronstrahlung m.b.H., Albert-Einstein Strasse 15, D-12489 Berlin, Germany}
\address{$^{4}$Central Research Institute of Electric Power Industry, Komae, Tokyo 201-8511, Japan}
\address{$^{5}$Institute of Scientific and Industrial Research, Osaka University, Ibaraki, Osaka 567-0047, Japan}

\begin{abstract}
The $ab$-plane optical conductivity  of eleven single crystals, belonging to the families  Sr$_{2-x}$CuO$_2$Cl$_2$, Y$_{1-x}$Ca$_{x}$Ba$_2$Cu$_3$O$_6$, Bi$_2$Sr$_{2-x}$La$_x$CuO$_6$, and Bi$_2$Sr$_{2}$CaCu$_2$O$_8$ has been measured with hole concentrations $p$ between 0 and 0.18, and for 6 K $\leq T \leq$ 500 K to obtain an infrared picture of the $p,T$ phase diagram of the  Cu-O plane.  At extreme dilution ($p$ = 0.005), a narrow peak is observed at 1570 cm$^{-1}$ (195 meV), that we assign to a single-hole bound state. For increasing doping, that peak broadens into a far-infrared (FIR) band whose low-energy edge sets the insulating gap. The insulator-to-metal transition (IMT) occurs when the  softening of  the FIR band closes the gap thus evolving into a Drude term. In the metallic phase, a multi-band analysis identifies a mid-infrared band which weakly depends on temperature and softens for increasing $p$, while the extended-Drude analysis leads to an optical scattering rate larger than the frequency, as found in other cuprates. The infrared spectral weight $W(T)$ is consistent with a Fermi liquid renormalized by strong correlations, provided that  the $T^4$ term of the Sommerfeld expansion is included above 300 K. In the superconducting phase, the optical response of single-layer Bi$_2$Sr$_{2-x}$La$_x$CuO$_6$ at optimum doping is similar to that of  the corresponding optimally-doped bilayer Bi$_2$Sr$_{2}$CaCu$_2$O$_8$.
\end{abstract}

\maketitle


\section{Introduction}

Despite twentyfive-year-long experimental and theoretical efforts, the low-energy electrodynamics of the high-$T_c$ cuprates still presents  phenomena, both in the normal and in the superconducting phase, which are not fully understood \cite{Timusk-99,Basov-05}.  Among them we can cite the pseudogap, which in the $ab$ plane does not manifest itself in the optical conductivity but only in the scattering rate, the bosonic spectral feature at 41 meV, the anomalous dependence of the spectral weight on frequency and temperature, the effect of the superconducting transition up to energies larger than the optical gap by one-two orders of magnitude, and so on. The understanding of those intriguing phenomena can be improved by collecting further spectroscopic data on the different high-$T_c$ families, in order to enucleate the common features of the Cu-O plane electrodynamics at low energy. Here we present an extended  study  in the  infrared of the family Bi$_2$Sr$_{2-x}$La$_x$CuO$_6$ (BSLCO), on the track of those 
previously performed, for instance, on  La$_{2-x}$Sr$_{x}$CuO$_4$ (LSCO) \cite{Lucarelli-03,Lee-05}, YBa$_2$Cu$_3$O$_{6+\delta}$ (YBCO)  \cite{Lee-05}, and Pb$_x$Bi$_{2-x}$La$_y$Sr$_{2-y}$CuO$_{6+\delta}$  \cite{Heumen}.

BSLCO is, like LSCO, a single-layer cuprate where the hole doping can be finely controlled  \cite{Ando-99}: the hole concentration $p$ can be varied from
0.03 to 0.19 by decreasing $x$ from 1.0 to 0.2, according to a well known \cite{Ono-03} non-linear relation. Here also, the rather low critical temperature of BSLCO ($T_c^{max}\simeq$ 33 K) allows for studies of the normal phase down to low $T$. In addition, its crystals can be  easily cleaved along the $ab$ planes, so that spurious contributions to the measured reflectivity from the $c$ axis are automatically ruled out.  On the other hand, the Cu-O planes of BSLCO exhibit a slight corrugation reminiscent of the 1: 5 superstructure of Bi$_2$Sr$_{2}$Cu$_2$O$_8$ (BSCCO) which here becomes incommensurate and doping dependent \cite{Lobben}. In BSCCO the only effect of the superlattice at infrared wavelengths is a slight difference between the  reflectivity along the $a$ and the $b$ axis, especially above the plasma edge which separates the intraband from the interband range \cite{Quijada}. In Bi$_2$Sr$_{2-x}$La$_x$CuO$_6$ we expect therefore that the slight and incommensurate corrugation of the Cu-O plane will be overwhelmed by the much stronger doping dependence of the reflectivity.

To extend our study below $p = 0.03$, we have  also  measured single crystals of Sr$_{2-x}$CuO$_2$Cl$_2$ (SCOC) with $x\simeq$ 0 ($p<0.005$) and $x$ = 0.3 ($p=0.005$), and of Y$_{0.97}$Ca$_{0.03}$Ba$_2$Cu$_3$O$_6$ (YCBCO) ($p=0.015$). 
In the present work we use  BSLCO, SCOC, and YCBCO to identify common absorption features in the Cu-O plane, irrespective of the out-of-plane lattice structure, and to investigate how they change under a variation in the hole-doping $p$ of nearly two order of magnitude.
In order to compare the optical behavior of  BSLCO  at optimum doping with that of BSCCO, both  below $T_c$ and well above room temperature, we have also measured the reflectivity of a single crystal of the 2212 compound with $T_c$ = 93 K.

 
\section{Experiment}

\begin{figure}[t]
\begin{center}
\leavevmode
\epsfxsize=10cm \epsfbox {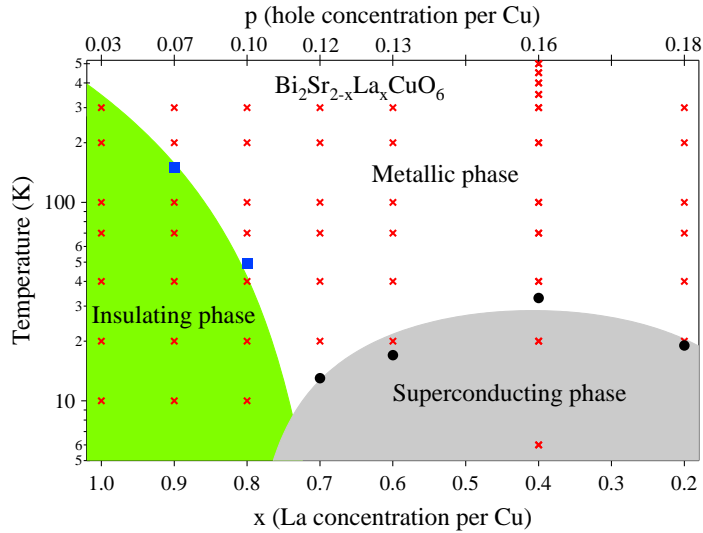}
\end{center}
\caption{Schematic $T,x$ and $T,p$ phase diagram of BSLCO from Ref. \cite{Ono-00} up to the miscibility limit. The crosses indicate the points where the present optical data were taken on the samples indicated in table 1. The dots mark the measured critical temperatures of the superconducting crystals, while square symbols refer to minima in the in-plane resistivity curves (see also figure 2).}
\label{PhaseDiagram}
\end{figure}

\begin{figure}[t]
\begin{center}
\leavevmode
\epsfxsize=9cm \epsfbox {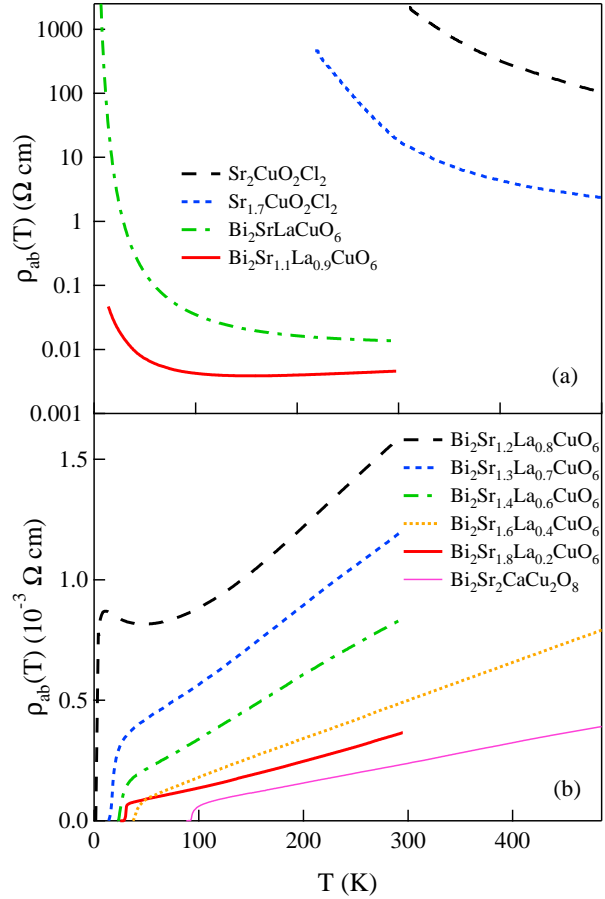}
\end{center}
\caption{Temperature dependence of the $ab$-plane resistivity of single crystals here studied: (a) insulators; (b) superconductors.}
\label{Rho}
\end{figure}

The SCOC,  BSLCO, and  BSCCO single crystals were all grown by the floating-zone technique, and characterized as described elsewhere \cite{Ono-03}. YCBCO was grown and characterized as described in Ref. \cite{Erb}.  The doping $x$, hole concentration $p$  of the samples, and the critical temperature $T_c$ of the superconductors,  are listed in table 1. The accuracy in the determination of $p$ for each $x$ value \cite{Ando-00} is $\pm10$\%.  $T_c$ was determined both from the zero resistivity and from the onset of the Meissner signal in the SQUID magnetization measurement, with an uncertainty of $\pm0.5$ K. For BSLCO, the temperatures and doping levels where the optical measurements were performed, are marked by the crosses in the ($x,T$) or ($p,T$) phase diagram of figure \ref{PhaseDiagram}. This reproduces that traced in Ref. \cite{Ono-00} on the basis of transport measurements in magnetic field, with slight adjustments of the coexistence lines.  Indeed, in figure \ref{PhaseDiagram} the IMT line (the $T_c$ line)  has been adapted to the squares (dots), which are extracted from the minima (the drop) in the $\rho_{ab}(T)$ curves of figure \ref{Rho}.








\begin{table}[h]
\begin{center}
\begin{tabular}{c|c|c|c}
\hline
\hline

Compound & $x$ & $p$ & $T_c$ (K) \\
\hline
\hline
Sr$_{2-x}$CuO$_2$Cl$_2$ & $\sim$0.0 & $<$0.005 & 0 \\
& 0.3 & 0.005 & 0 \\
\hline
Y$_{1-x}$Ca$_x$Ba$_2$Cu$_3$O$_6$ & 0.03 & 0.015 & 0 \\
\hline
Bi$_2$Sr$_{2-x}$La$_x$CuO$_{6}$ & 1.0 & 0.03 & 0 \\
& 0.9 & 0.07 & 0 \\
& 0.8 & 0.10 & 1.4 \\
& 0.7 & 0.12 & 13 \\
& 0.6 & 0.13 & 17 \\
& 0.4 & 0.16 & 33 \\
& 0.2 & 0.18 & 19 \\
\hline
Bi$_2$Sr$_{2}$CaCu$_2$O$_8$ & & 0.16 & 93 \\

\hline
\hline

\end{tabular}

\caption{Vacancy or dopant concentration $x$, hole doping per Cu ion $p$, and critical temperature $T_c$ for
the single crystals here measured in the infrared.}
\end{center}
\label{TableSamples}
\end{table}


Figure \ref{Rho} indeed shows the resistivity $\rho_{ab}(T)$ of both  optimally-doped samples,  Bi$_2$Sr$_{1.6}$La$_{0.4}$CuO$_{6}$ and Bi$_2$Sr$_2$CaCu$_2$O$_8$ up to 500 K, for the other crystals up to 300 K.  The superconductors in figure \ref{Rho}(b), at least in the optimal doping region, display the "famous" linear resistivity behavior from $T_c$ to the highest $T$. In  BSLCO with $x$ = 0.8 and $p$ = 0.10 instead the metallic behavior appears above 50 K only, while  a broad minimum is observed above its $T_c\simeq1.4$ K. 
The $\rho_{ab}(T)$ of the non-superconducting samples is reported in figure \ref{Rho}(a). It displays a typical insulating behavior ($\mathrm{d}\rho_{ab}/\mathrm{d}T<0$) for all of them, except for BSLCO with $x$ = 0.9 $p\simeq0.07$, which looks like a poor metal above $\sim$ 100 K. Below 50 K, its $\rho_{ab}(T)$ diverges for $T\rightarrow0$ according to a variable range hopping regime \cite{Ono-03}. 
A check value of $\rho_{ab}$ was determined  also for the Y$_{0.97}$Ca$_{0.03}$Ba$_2$Cu$_3$O$_6$  at 200 K \cite{Janossy-private}. The very large resistivity of  the SCOC insulators has been reported at high temperature only, where the measuring range of our apparatus allowed for reliable measurements.

Basing on the data of figure \ref{Rho}, the IMT of the Cu-O plane in BSLCO can be placed between $p = 0.07$ and $p = 0.10$. This finding is consistent with the Ioffe-Regel limit, which fixes the insulator-to-metal crossover where the mean free path $\ell$ is equal to the Fermi wavelength $\lambda_F=2\pi/k_F$. Apart from numerical factors,
this leads for the Cu-O plane of both BSLCO and LSCO \cite{Ono-00} to the condition \cite{Ando-08}
$k_F\ell=(hc_0/e^2\rho_{ab})\sim1$. Here $c_0$ is the $c$-axis lattice
spacing. Indeed, following Ref. \cite{Ono-00}, from the $\rho_{ab}$ (10 K)
in figure \ref{Rho}, one obtains $k_F\ell=3.4$ at $p = 0.10$ and $k_F\ell=0.05$
at $p = 0.07$. Magnetic fields on the order of 60 T
displace the IMT \cite{Ono-00} to $p\simeq1/8$.

The $ab$-plane reflectivity $R(\omega)$ of the single crystals, obtained from the same batch of those used for resistivity measurements, has been measured at near-normal incidence with Michelson interferometers. The dimensions of the $ab$ surfaces were between $1.5\times1$ mm$^2$ and $8\times3$ mm$^2$. Their thickness along the $c$ axis was $\sim50$ $\mu$m for all Bi-based crystals, 0.5 mm for YCBCO, 150 (180) $\mu$m for Sr$_{2-x}$CuO$_2$Cl$_2$ with $x\simeq$ 0 ($x$ = 0.3). 

The measurements were performed after thermoregulating the samples within $\pm$ 1 K  between 10 or 40 K and $\pm$ 3 K between 300  and 500 K. In BSLCO and SCOC,  they were made shortly after cleaving the sample surfaces parallel to the Cu-O planes, thus excluding any contribution from the $c$ axis.
The reference in the infrared (visible) range was a gold (silver) film evaporated \textit{in situ} onto
the sample. This  was mounted in a closed-cycle cryostat or,
above $T=300$ K,  inside an optical  chamber under vacuum. 
At every $T$, and for any spectral range, the intensity reflected by the sample was measured with and without coating, in order to compensate for the displacements of the sample holder due to thermal expansion. The chemical stability of both samples was checked by measuring $R(\omega)$ at 300 K after every high-$T$ cycle.

In order to investigate the superconducting phase in the $ab$-plane reflectivity of the optimally doped samples, due to the small gap related to the relatively low $T_c$'s of Bi$_2$Sr$_{1.6}$La$_{0.4}$CuO$_{6}$, it was necessary to extend our measuring range to the sub-Terahertz region (1 THz $\simeq33$ cm$^{-1}$). The necessary brilliance was provided by the Coherent Synchrotron Radiation (CSR) extracted from the storage ring BESSY II, working in the so-called \emph{low-$\alpha$} mode \cite{AboBakr-03}.  Below 25 cm$^{-1}$ we could thus measure the absolute reflectivity with a sensitivity of 1 \% \cite{CaAlSi}. From 20 to 40 cm$^{-1}$ we used incoherent synchrotron radiation and above that frequency conventional sources.
The real part  $\sigma_1(\omega)$ of the $ab$-plane optical conductivity was obtained from $R(\omega)$ through Kramers-Kronig (KK) transformations. Preliminary extrapolations of $R(\omega)$ to $\omega$ = 0  were based on Drude-Lorentz fits, which provided a $\sigma_1(0)$ which deviated from the  $\sigma_{dc}$  measured at the same $T$ within a few percent. Afterwards, these fits were adjusted exactly to $\sigma_{dc}$.
In the superconducting phase the extrapolation was based on the London formula $1-R(\omega)\propto\omega^{-4}$. The extrapolations to high frequency were based instead on the data of Ref. \cite{Terasaki-90} up to 12.5 eV (100,000 cm$^{-1}$) and on a power law beyond this energy.


\section{The insulating phase and the insulator-to-metal transition}

\begin{figure}[t]
\begin{center}
\leavevmode    
\epsfxsize=8.5cm \epsfbox {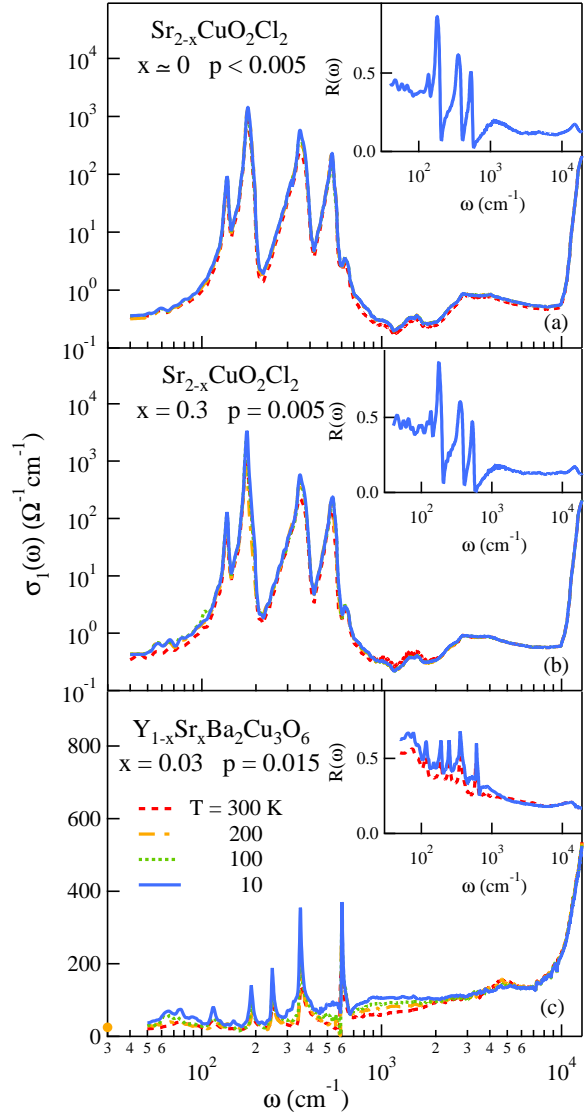}
\end{center}
\caption{Optical conductivity of the three lowest-doping single crystals from the far-IR to the visible, belonging to different cuprate families. In panel (c) the circle on the vertical axis mark the dc conductivity at 200 K. In the insets, the reflectivity is reported in the same spectral range at 300 and 10 K.}
\label{sigma-insul}
\end{figure}

\begin{figure}[t]
\begin{center}
\leavevmode    
 \epsfxsize=8.5cm \epsfbox {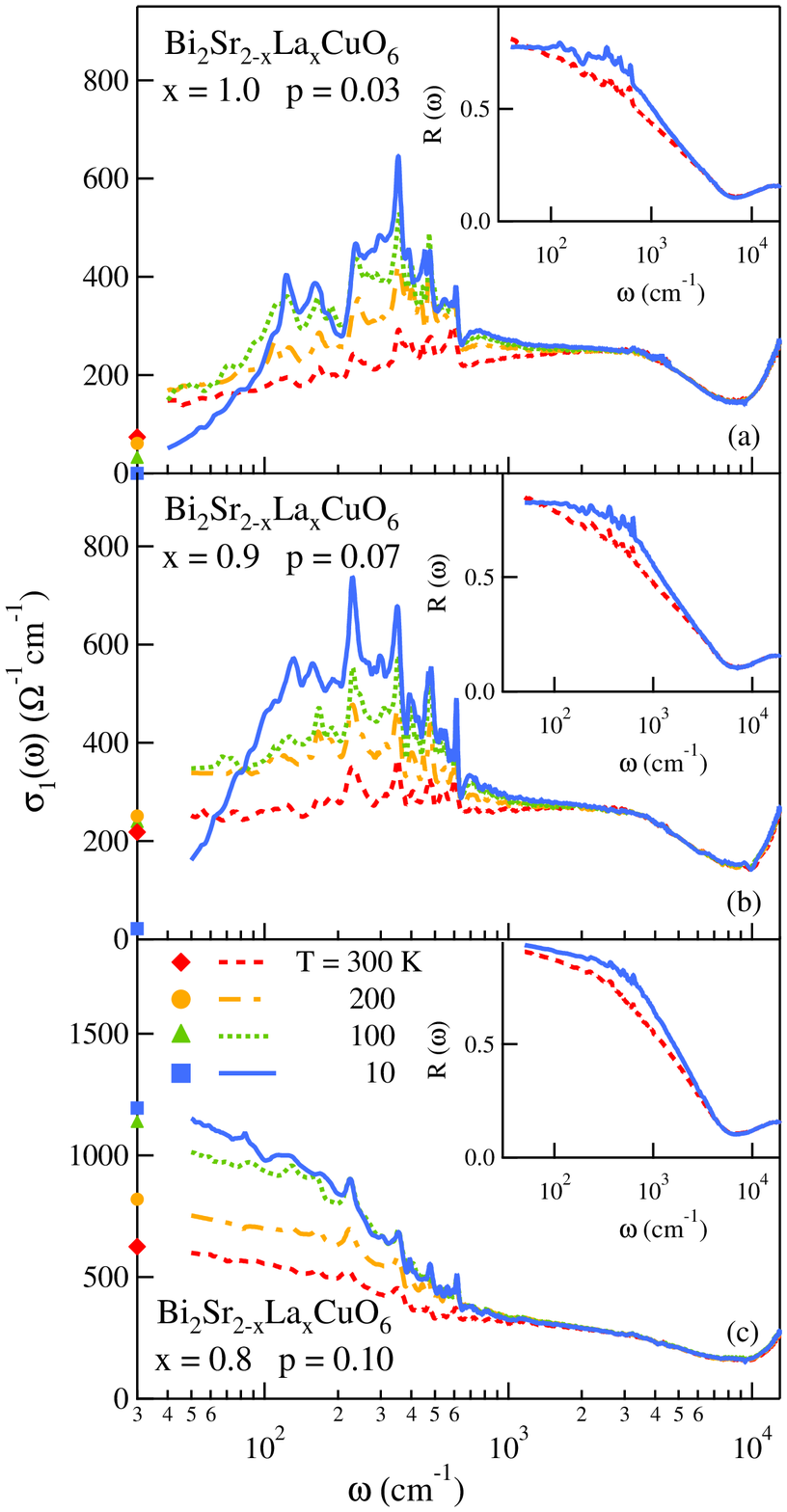}
\end{center}
\caption{Optical conductivity of  crystals with increasing hole doping across the insulator-to-metal-transition of the Cu-O plane. The diamond symbols on the vertical axis mark the dc conductivity measured on samples of the same batch. In the insets, the reflectivity  is reported in the same spectral range at 300  and 10 K.}
\label{sigma-MIT}
\end{figure}

The optical conductivity of the lowest-doping insulators belonging to each family are reported in figure  \ref{sigma-insul} for SCOC and YCBCO, and in figure   \ref{sigma-MIT}(a) for BLSCO. The far infrared spectra are dominated by the peaks of the transverse optical (TO) phonons. At higher frequencies a broad contribution (mid-infrared, or MIR band) appears, already observed in many strongly correlated materials \cite{Paolone}-\cite{Basov2011}. Its physical interpretation will be discussed in the next section.
The edge of the charge transfer (CT) excitation from the O $2p$ to the Cu $3d_{x^2-y^2}$ orbitals can be instead observed in the visible. 
The peaks of the TO, infrared-active $E_u$ optical phonons of the $ab$ plane are four in SCOC and six in YCBCO. In BSLCO one observes five pairs of lines instead of  the six predicted by a factor-group analysis. Indeed, the bending mode involving the apical oxigen cannot be detected because of its low intensity \cite{Tajima-91}. The splittings of the remaining lines were already reported and explained in terms of  distorted double Bi-O layers \cite{Tajima-91}.  The lowest-frequency phonons can be assigned to translational modes of rare earth Bi and Sr(La) translational modes, while the mode at 445-480 cm$^{-1}$ is assigned to the Bi-O bending.
On the other hand, the phonon doublets at 350-390 cm$^{-1}$ and 560-600 cm$^{-1}$ involve only Cu and the O atoms building up the ochtaedral structures. They  are associated with bending and stretching modes in Cu-O planes, respectively. The frequencies of all lines observed at  the lowest measured $T$, as obtained by fits to Lorentzians, are listed in table 1 and compared with those reported at 300 K in Ref. \cite{Tajima-91}. The phonon frequencies of YCBCO are very similar to 
those previously measured \cite{Tajima-91} on pure YBa$_2$Cu$_3$O$_y$.


The $ab$-plane optical conductivity $\sigma_1(\omega)$ of three single crystals of Bi$_2$Sr$_{1-x}$La$_{x}$CuO$_{6}$ with $x$ = 1.0, 0.9, and 0.8 ($p$ = 0.03, 0.07, and 0.10, respectively) is shown in
figure \ref{sigma-MIT} at selected temperatures in the normal state. The corresponding reflectivity spectra are shown in the insets. The  $\sigma_{dc}$ values measured on samples belonging to the same batch and at the same temperatures are also shown. Together with those of figure \ref{sigma-insul}, the curves of figure \ref{sigma-MIT}  monitor the evolution of  $\sigma_1(\omega)$ with $p$ and $T$ as the doped Cu-O plane approaches, and eventually overcomes, the IMT at $p \sim 0.08$.


\begin{table}[h]

\begin{center}
\begin{tabular}{{l}{c}{c}{c}{c}{c}{c}{c}{c}}
\hline
 \hline

	Sample			& $E_u$(1) & $E_u$(2) &	$E_u$(3)	& $E_u$(4) & $E_u$(5) & $E_u$(6) 	\\
  \hline
 
Sr$_{2}$CuO$_2$Cl$_2$  							&	137	&	179	&	350	&	526	&	  &	\\
Sr$_{2}$CuO$_2$Cl$_2$ \cite{Tajima-91}				&	140	&	176	&	351	&	525	&	 &	\\
Y$_{0.97}$Ca$_{0.03}$Ba$_2$Cu$_3$O$_6$  			&	   & 116	&	189	&	249	&	355	& 605	\\
YBa$_2$Cu$_3$O$_{6.1}$ \cite{Tajima-91} 			&	&  118	&	193	&	250	&	357	& 588	\\
Bi$_2$Sr$_{1.0}$La$_{1.0}$CuO$_{6}$  				&	121, 163\	 & 238, 277 \ & 352, 390\   & 448, 477\  &  	 & 553, 608	\\
Bi$_2$Sr$_{1.0}$La$_{1.0}$CuO$_{6.5}$ \cite{Tajima-91} 	& 130, 170\ & 230, 260\ & 350, 390\ & 445, 480\ &  	&	560, 600\	\\
\hline
\hline
\end{tabular}
\end{center}
\label{modes}
\caption{Transverse optical phonon frequencies in cm$^{-1}$, measured with a resolution of 2 cm$^{-1}$ at the lowest temperature of the experiment (40 K for SCOC, 10 K for YCBCO and BSLCO) in the $ab$-plane of different crystals at their lowest doping, compared with data in the literature at 300 K. In BSLCO, all doubly degenerate $E_u$ modes are split by a distortion of the Bi-O layers, while a bending mode of the apical oxygen ($E_u$(5)) is not detected}.
\end{table}


\begin{figure}[t]
\begin{center}
\leavevmode    
\epsfxsize=12cm \epsfbox {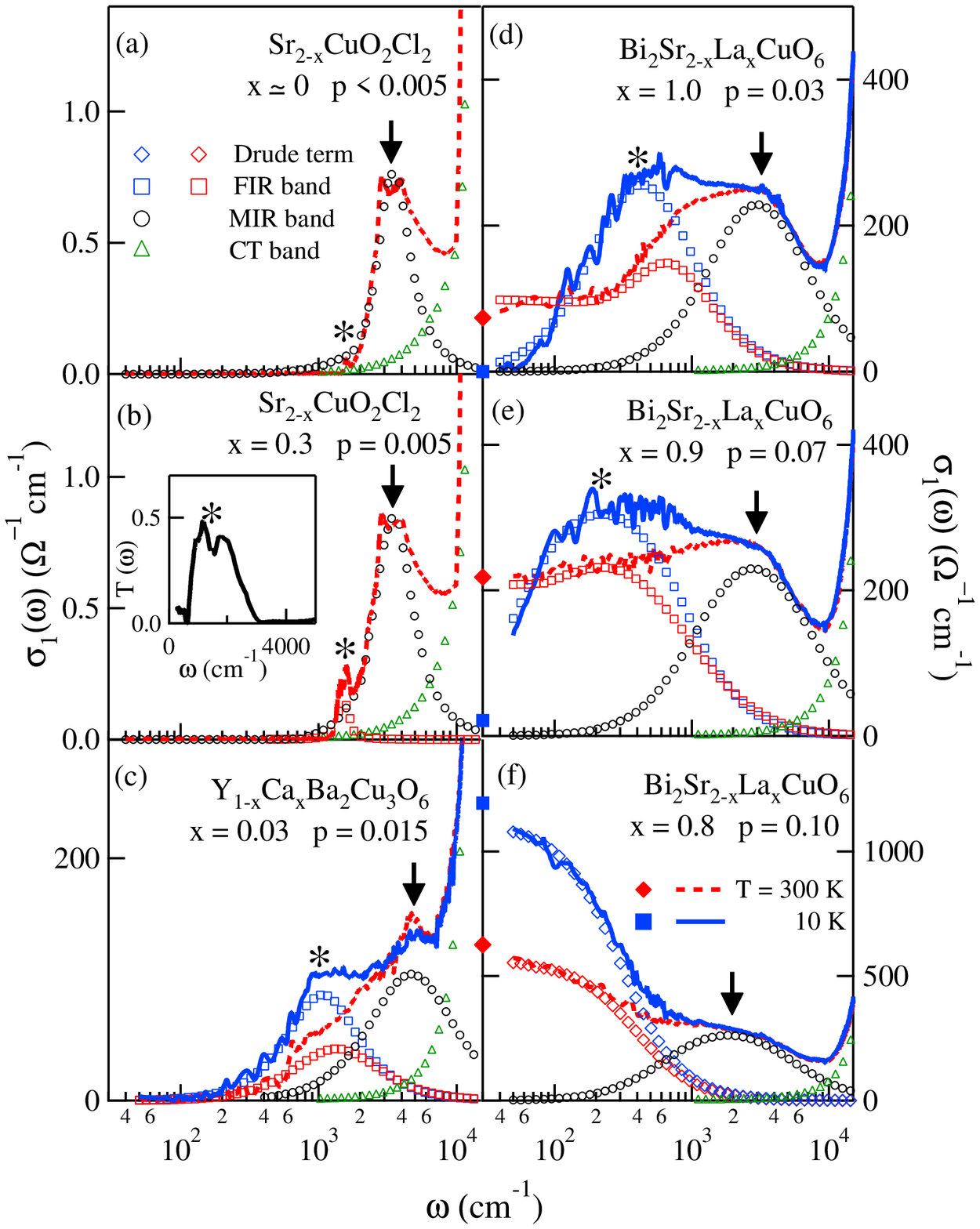}
\end{center}
\caption{Optical conductivity, after subtraction of the phonon peaks via
Lorentzian fits of six crystals of SCOC, YCBCO and BSLCO whose properties are described in Table II. Their full $\sigma_1(\omega)$  are shown in figures \ref{sigma-insul} and \ref{sigma-MIT}.  The remaining contributions are indicated by open symbols, as reported in panel (a). The stars (arrows) mark the peak frequencies $\omega_{FIR}$ ($\omega_{MIR}$) (see text). In panel (b)  the transmittance of the SCOC (p=0.005) sample is also reported, showing an absorption at $E_0$ = 1570 cm$^{-1}$. The $\sigma_{dc}$ values on the vertical axis show that  the extrapolations to $\omega = 0$ are not affected by the phonon-line subtraction.} 
\label{PhonSub}
\end{figure}

In order to better follow that evolution, in figure \ref{PhonSub} we show the $ab$-plane conductivity of the six crystals with the lowest non-zero doping after subtracting all phonon contributions by fits similar to those whose results are reported in table 2. The remaining contributions to $\sigma_1(\omega)$, peaked at zero frequency (Drude term), and at a far-infrared (mid-infrared) frequency  $\omega_{FIR}$ ($\omega_{MIR}$), are indicated by open symbols. 

In figure \ref{PhonSub} the effect of increasing doping on the Cu-O plane clearly appears. 
At $p<0.005$ (SCOC sample in (a)) mid-IR absorption can be observed, which includes a weak peak (marked by a star) at $E_0$ = 1570 cm$^{-1}$ (195 meV) and a broad structured band which extends between 3000 and 4000 cm$^{-1}$. A linear extrapolation of the absorption to zero conductivity places its edge at $E_g \sim$ 2000 cm$^{-1}$ (0.25 eV). A mid-IR absorption has been observed in several highly correlated materials and associated with different excitation mechanisms \cite{Paolone}-\cite{Basov2011}. In the SCOC material, the mid-IR broad band was previously ascribed to a Frenkel exciton, probably of crystal-field origin, associated with strong multimagnon sidebands \cite{Perkins, Zibold}. However, as the charge density in the $x\simeq$ 0 antiferromagnetic SCOC sample is nearly zero, theoretical calculations for a single hole in the AFM Cu-O plane suggest that this band is related to the photoionization process of an isolated hole strongly interacting with the magnetic degrees of freedom (magnetic polaron) \cite{Mishchenko-08}. The narrow peak, instead, was assigned either to a phononic hypertone in Ref.\cite{Perkins} or, in Ref.\cite{Zibold}, to an extrinsic absorbtion due to chlorine surface contamination.   
Here, the fine tuning of doping in our SCOC crystals allows us to track the evolution of the mid-IR absorption adding a controlled number of holes in the Cu-O plane. In particular, for the SCOC crystal with $x$ = 0.3 and $p$ = 0.005 (figure \ref{PhonSub}(b)) both the structured mid-IR band and the narrow peak increase their intensity, while mantaining the same characteristic frequencies of those for $x\simeq$ 0. This allows us to propose an alternative interpretation for both bands, namely that they have an intrinsic origin related to the added holes. Indeed, the observation that the optical conductivity in figure \ref{PhonSub}(b) is typical of semiconductors at very low doping, reasonably allows one to associate the feature at $E_0$ with a hole bound to the Sr vacancies.  According to the usual hydrogen-like model, the ground state of the bounded hole is 

\begin{equation}
E_{0}=\frac{m_b/m}{\epsilon_r^2} \cal R
\label{E0}
\end{equation}

\noindent
where $\cal R$ =109737 cm$^{-1}$ is the Rydberg constant, $m_b$ the hole effective mass, $m$ the free electron mass, and $\epsilon_r$ the relative dielectric constant  of the insulating cuprate. If one assumes $m_b=m$, for  $E_0$ = 1570 cm$^{-1}$ one obtains $\epsilon_r$ = 8.4, a value consistent with Ref.  \cite{Mark}.
Thanks to the SCOC samples, we can thus correct the $E_0 \simeq$ 1000 cm$^{-1}$ reported in a previous paper  \cite{Lupi-09},  where  it was taken from the gap edge at the minimum doping then available, namely $p$ = 0.015.

\begin{figure}[t]
\begin{center}
\leavevmode    
\epsfxsize=9.5cm \epsfbox {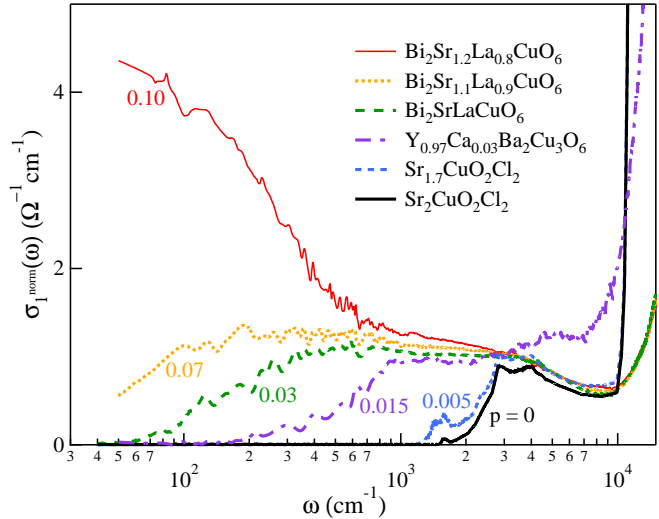}
\end{center}
\caption{Same $\sigma_1(\omega)$ as in figure \ref{PhonSub} at the lowest $T$ of the experiment, but normalized to 1 $\Omega^{-1}$ cm$^{-1}$ around 3000 cm$^{-1}$, to show how  the insulating gap is filled up by increasing doping.}
\label{IMT-gap}
\end{figure}

As $p$ further increases below $p_{IMT}$, the weak feature detected at $p$ = 0.005 appears to turn into a broad band, whose peak at $\omega_{FIR}$ (marked by stars in figure \ref{PhonSub}) softens rapidly with doping and whose edge progressively closes the insulating gap, as better shown pictorially in figure \ref{IMT-gap}. Therein, $\sigma_1^{norm}$ is the  same $\sigma_1$ of the samples in figure \ref{PhonSub} at the lowest $T$, but  normalized for all of them to 1 $\Omega^{-1}$ cm$^{-1}$ around 3000 cm$^{-1}$.
The MIR band at $\omega_{MIR}$ in figure \ref{PhonSub} also softens, even if much less dramatically. Already at $p$ = 0.03, the gap survives at low temperature only, while at high $T$ it is replaced by a broad, non-Drude like absorption which may be ascribed to incoherent charge hopping. At $p$ = 0.07 the gap is closed also at the lowest $T$, but the dc conductivity is still very small. A full metallic phase, with a Drude-like $\sigma_1 (\omega)$ at all temperatures, is finally established at $p$ = 0.10. It is then confirmed by the present optical data that the IMT in the Cu-O plane can be placed between 0.07 and 0.10. holes per copper site. The behavior with $p$ of the gap in figure \ref{IMT-gap} is consistent with that measured by ARPES in BSLCO at the leading edge midpoint along the nodal line of the Brillouin zone \cite{Shen}. 

As discussed in a previous paper \cite{Lupi-09}, the IMT can be explained qualitatively by a Mott mechanism, which is further supported by the present observation of a hole bound state in the insulating gap at extreme dilution. We can thus make a prediction on the critical hole density $p_{IMT}$.  Indeed,  the radius of the hole orbit is related to the experimental $E_0$ by

\begin{equation}
R^2= \frac{3}{2}a_0^2\frac{\cal R}{E_0}
\label{E0}
\end{equation}

\noindent
where $a_0$  is the Bohr radius. One thus obtains $R$= 0.54 nm. The Mott  condition \cite{nota}  implies that the hole density per unit area at the transition

\begin{equation}
\rho_{IMT}\sim\frac{1}{\pi R^2}
\label{rho}
\end{equation}

and the Cu density $\rho_{\mathrm{Cu}}=1/a^2$ satisfies the relation

\begin{equation}
p_{IMT}=\rho_{IMT}/\rho_{\mathrm{Cu}}\sim\frac{a^2}{\pi R^2}
\label{pMIT}
\end{equation}

\noindent
In SCOC $a$ = 0.39716 nm \cite{Miller}. By using the result of equation \ref{E0} one obtains
$p_{IMT}\simeq 0.17$, a value larger by a factor of two than the experimental $p_{IMT}$ observed in figure \ref{PhonSub}. This  should not be seen as unsatisfactory, if one considers that in a much simpler semiconductor like Si:P  the difference between the observed  $p_{IMT}$ and that predicted by the Mott criterion is much larger \cite{Kittel}. Here, the main reason for the disagreement may be the fact that $R \sim a$ makes unreliable the approximation of the Cu-O plane as an uniform dielectric.

The broad  contributions peaked at  $\omega_{FIR}$ and $\omega_{MIR}$  in figure \ref{PhonSub} have been observed in a number of  cuprates at low doping \cite{Basov-05},\cite{Calvani-96}-\cite{Perucchi-09} and their origin has been long discussed  in the literature. 
For the FIR band, a possible explanation is disorder. In cuprates added with Zn impurities or irradiated
by high-energy particles, the metallic phase can be
destroyed and the Drude spectral weight strongly reduced,
due to a poor screening of the impurities and to the resulting
fluctuating potentials in the Cu-O planes \cite{Basov-98}.
Recent calculations \cite{Atkinson-02} show that such disorder effects
are amplified in a $d$-wave electronic symmetry and that,
for increasing impurity content, the Drude term turns into a
FIR peak at a finite frequency. However,  in the present experiment the gap is monotonically closed, and the metal approached, both by  adding  (in SCOC and YCBCO) and by subtracting impurities (in BSLCO). Therefore, the parameter which governs the metal-to-insulator
transition is $p$ rather than $x$.
The above scenario is also consistent with recent calculations of the optical conductivity in hole-doped cuprates. They are based on a $t$-$J$-Holstein approach, where the FIR band has a dominant electron-phonon character (lattice polaron), while the MIR band is attributed mainly to the electron-spin interaction (spin polaron) \cite{Mishchenko-08}.
This polaronic scenario seems  to be also consistent with other experimental observations. 
Indeed, in BSLCO the FIR peak behaves with $p$ like the FIR band of electron-doped NCCO vs the electron concentration $n$
 \cite{Lupi-99}. This absorption was attributed to large polarons
\cite{Calvani-01}, and its softening was explained in terms of
polaron-polaron interactions which increase with $n$ \cite{Tempere-01,Lorenzana-01}.
At room temperature, where $k_BT\approx\Delta$,
incoherent polaron hopping takes place: this may explain
the flat background observed in figures \ref{PhonSub}(d) and \ref{PhonSub}(e) and the
resulting, nonvanishing $\sigma_{dc}$ at $p<p_{IMT}$. 

As $p$ increases, $\omega_{MIR}$ in figure \ref{PhonSub} shifts steadily to lower energies, to reach $\sim$ 4500 cm$^{-1}$ at $p  = 0.015$. This  value is consistent with the determination of the MIR peak in the other cuprates ($\sim$ 0.5 eV) \cite{Basov-05}.

\section{The metallic phase}
\subsection{The optical conductivity}

\begin{figure}[t]
\begin{center}
\leavevmode    
\epsfxsize=8.5cm \epsfbox {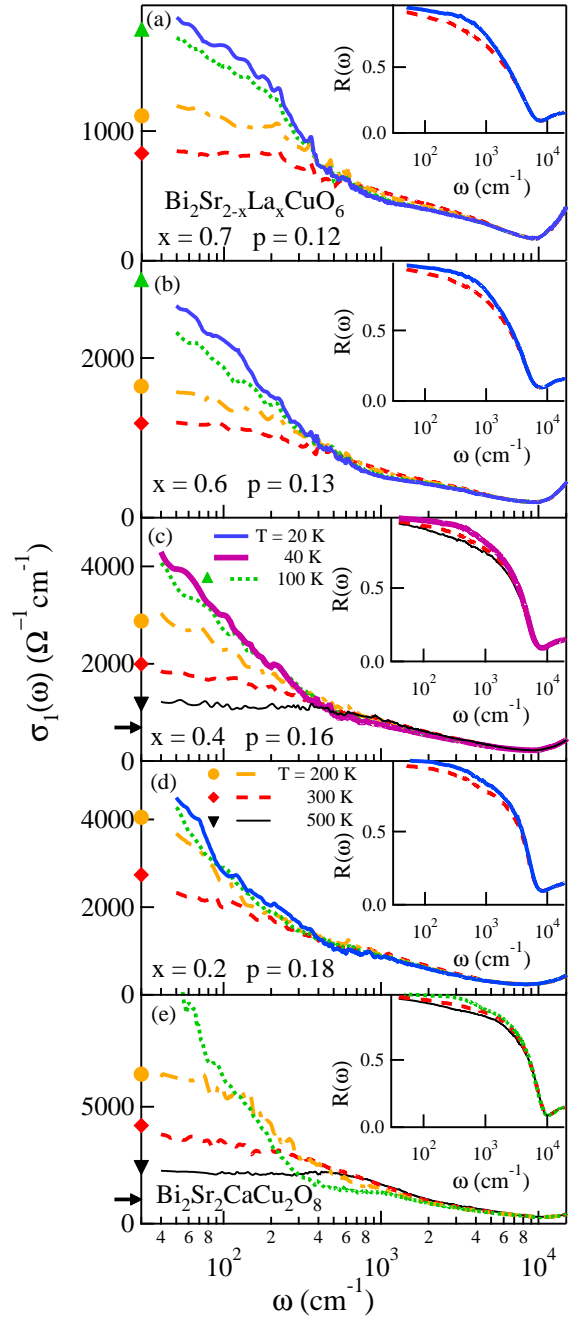}
\end{center}
\caption{Normal-phase optical conductivity of
Bi$_2$Sr$_{2-x}$La$_x$CuO$_6$ with different La concentrations, and of Bi$_2$Sr$_{2}$CaCu$_2$O$_8$, as obtained from the reflectivity spectra in the corresponding insets.
The symbols on the vertical axis indicate the dc conductivity measured on samples of the same batch ($\sigma_{dc}$ values at the lowest $T$'s are out of scale). The horizontal arrows in panel (c) and (e) mark the Ioffe-Regel limit for the dc conductivity of optimally doped BSLCO and BSCCO (see text).}
\label{sigma-metal}
\end{figure}


Figure \ref{sigma-metal} shows the $ab$-plane reflectivity and the optical conductivity of four metallic BSLCO samples in the normal phase, from the underdoped to the overdoped regime, and those of the optimally doped BSCCO crystal for comparison. 
The $R(\omega)$ in all the insets has a minimum at $\omega\simeq$ 8000 cm$^{-1}$, which is usually assumed to approximately separate the intraband absorption from the interband absorption.  

\begin{figure}[t]
\begin{center}
\leavevmode    
\epsfxsize=9.5cm \epsfbox {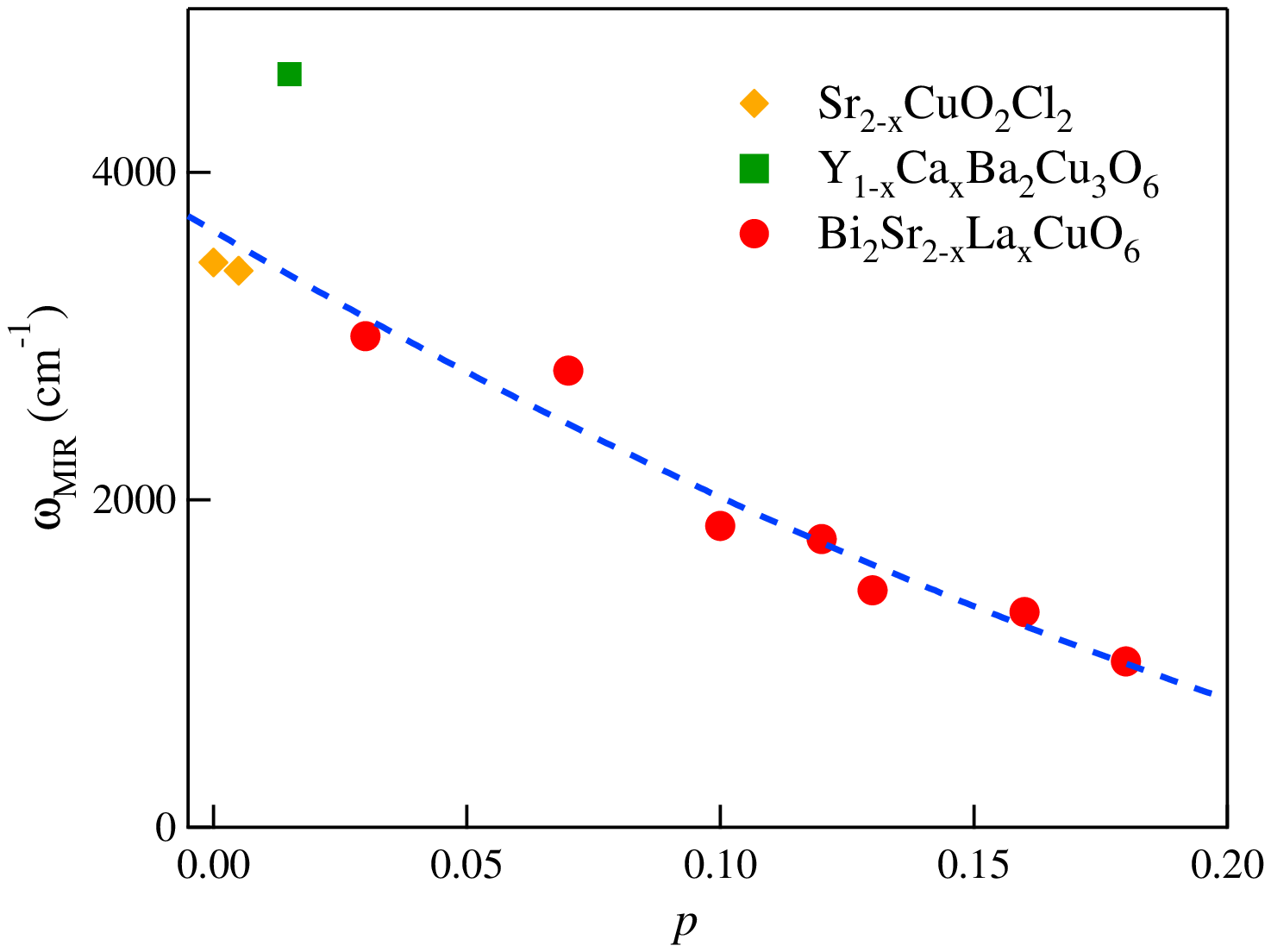}
\end{center}
\caption{Peak frequency (independent of $T$) of the MIR band vs. hole concentration in the Cu-O plane. Data are provided by FIR+MIR or Drude+MIR fits to the  $\sigma_1(\omega)$ in figures \ref{PhonSub} and \ref{sigma-metal}. The line is a guide to the eye.}
\label{omega_MIR}
\end{figure}

As usually done in cuprates, the optical conductivity of Fig. \ref{sigma-metal} can be analyzed in two different ways. In the two-component model, the former oscillator is a standard  Drude peak centered at $\omega=0$, which increases in intensity with $p$ and broadens as $T$ increases. The second component is a mid-IR band peaked at $\omega_{MIR}$ and described in terms of a Lorentzian shape, like in the less doped crystals in figure \ref{PhonSub}. Indeed, that band survives at $p > p_{IMT}$ as shown clearly  in figure \ref{IMT-gap}, where  the $\sigma_1(\omega)$ of the underdoped crystal with $x$ = 0.8 and $p$ = 0.10 is shown after phonon subtraction. Here, a clear change of slope is found above 1000 cm$^{-1}$ and the two-component fit provides $\omega_{MIR} \simeq$ 1800 cm$^{-1}$. As the Drude intensity increases with doping, for $p >$ 0.10 the MIR band is no more resolved. However,  an additional oscillator in the mid infrared is required by all Drude-Lorentz fits to the spectra in figure \ref{sigma-metal}. The MIR band continues to shift remarkably toward lower energies as $p$ increases, as shown in figure \ref{omega_MIR}. At any doping the MIR band is instead nearly insensitive to  temperature between 10 and 300 K.

\begin{figure}[t]
\begin{center}
\leavevmode    
\epsfxsize=9cm \epsfbox {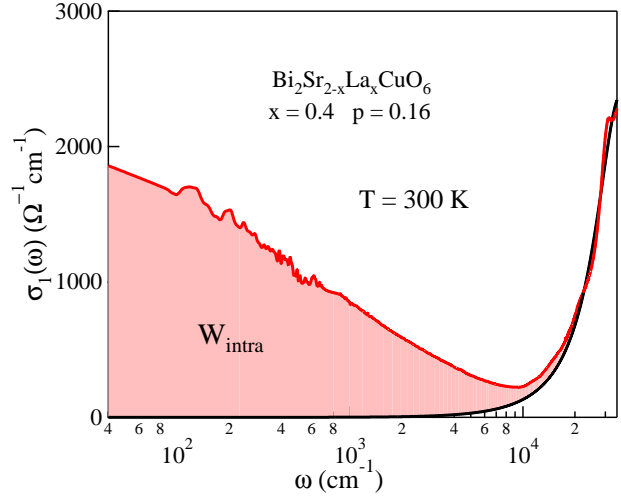}
\end{center}
\caption{Example of extraction of the unscreened plasma frequency from the $\sigma_1(\omega)$ in figure \ref{sigma-metal} for BSLCO with $x$ = 0.4, via Lorentzian fits to the intraband (Drude+MIR) and interband absorption and subsequent evaluation of the intraband spectral weight (colored area).}
\label{subtraction}
\end{figure}

The fact that a band interpreted in terms of magnetic polarons \cite{Mishchenko-08} survives in the metallic phase of a cuprate family is not unexpected. Indeed, the existence of local antiferromagnetic fluctuations has been reported for example in YBCO  \cite{Lee-05}, and spin-density waves are known to exist in superconducting pnictides \cite{Perucchi-09}.  The data of figure \ref{omega_MIR} are particularly interesting if one considers the possible role of magnetic excitations in the pairing mechanism. For example,  according to the resonant-valence-bond (RVB) theory \cite{Anderson} the pairing is intimately related to superexchange interaction of energy $J \sim t^2/U$, where $t$ is the effective hopping matrix element and $U$ is the Hubbard on-site repulsion. Other approaches to high-$T_c$ superconductivity based on the $t-J$ model also involve the $J$ scale of energy \cite{Maier}: it would be interesting to investigate theoretically, in those frameworks, the softening of $\omega_{MIR}$ with $p$. 

\begin{figure}[t]
\begin{center}
\leavevmode    
\epsfxsize=12cm \epsfbox {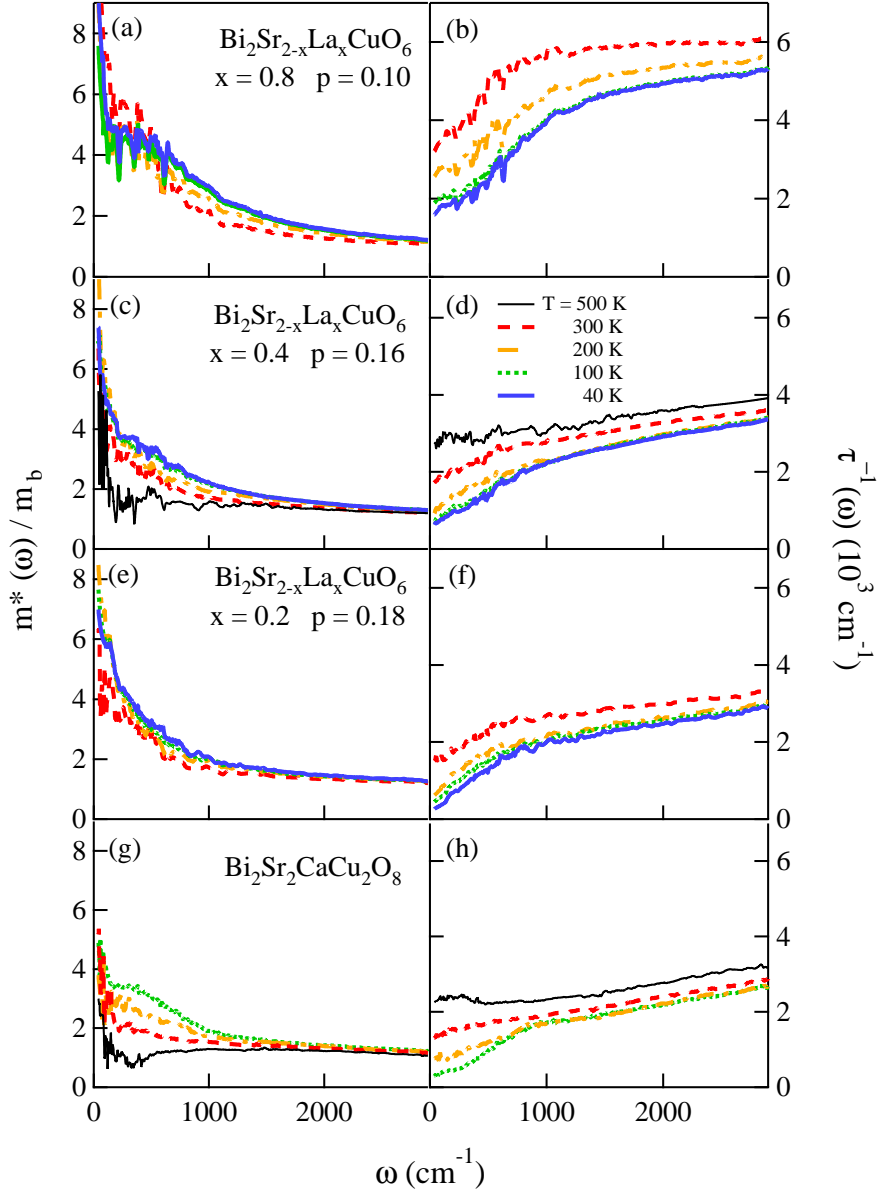}
\end{center}
\caption{Renormalized carrier mass $m^*(\omega)/m_b$ (left panels) and frequency-dependent scattering rate $\tau^{-1}(\omega)$ (right panels) as extracted from the $\tilde \sigma(\omega)$ of  the $ab$ plane in metallic BSLCO and in BSCCO  through the extended Drude model.}
\label{S-ExtDrude}
\end{figure}

An alternative picture for the metallic optical conductivity of cuprates is the one-component \emph{extended Drude model}. Here one describes the complex   $\tilde \sigma(\omega)$  in the whole infrared range by a single contribution peaked at $\omega$ = 0, weighted by the unscreened plasma frequency $\omega_p$ but with an optical scattering rate $\tau^{-1}(\omega)$ and an effective mass $m^*(\omega)$
which are both dependent on frequency:

\begin{equation}
\tilde{\sigma}(\omega)=\frac{1}{4\pi}\frac{\omega_p^2}{\tau^{-1}(\omega)-i\omega m^*(\omega)/m_b}
\label{Anomalous}
\end{equation}

\noindent
Therefore, in terms of the complex dielectric function, one has \cite{Heumen07}  
 
\begin{equation}
\tau^{-1}(\omega) = \frac{\omega_p^2}{\omega} \frac{\epsilon_2(\omega)}{[\epsilon_{\infty} - \epsilon_1(\omega)]^2 + \epsilon_2^2(\omega)}
\label{tau}
\end{equation}

and

\begin{equation}
\frac{m^*(\omega)}{m_b} =  \frac{\omega_p^2}{\omega^2} \frac{\epsilon_{\infty} - \epsilon_1(\omega)}{[\epsilon_{\infty} - \epsilon_1(\omega)]^2 + \epsilon_2^2(\omega)}
\label{m*}
\end{equation}

Here the unscreened plasma frequency $\omega_p$ was extracted from the $\sigma_1(\omega)$ in figure \ref{sigma-metal} via Lorentzian fits to the intraband (Drude+MIR) and interband absorptions. This allows one to subtract the interband contribution and  then estimate the intraband spectral weight $W_{intra}$. Indeed, in the units here employed, $\omega_p = \sqrt{(120/\pi) W_{intra}}$ \cite{Hwang-07}. We found out $\omega_p$ $\simeq$ 11000, 14000, 14500, and 17500 cm$^{-1}$ for BSLCO with $x$ = 0.8, 0.4, 0.2, and BSCCO, respectively.
An example of fit is shown in figure \ref{subtraction} for BSLCO with $x$ = 0.4. Moreover, $\epsilon_{\infty}$  was obtained by subtracting the intraband contribution to $\sigma_1(\omega)$, recalculating the remaining $\epsilon_1(\omega)$, and extrapolating the result to zero frequency. We obtained $\epsilon_{\infty}\simeq$4.5 (independent of the material), a standard value for the cuprates \cite{VdMNature}. The final results for  $m^*(\omega)$ (left panels) and $\tau^{-1}(\omega)$ (right panels) are shown in figure \ref{S-ExtDrude} for 
underdoped (a,b), optimally doped (c,d), and overdoped (e,f) BSLCO crystals, as well as for the optimally-doped BSCCO crystal (g,h). One sees that $m^*(\omega)$ approaches the band mass  $m_b$ at high energy, while in the far infrared it reaches values  $\sim 8 m_b$. Moreover, $m^*(\omega)$ is just slightly affected by the change in the hole content, consistently with previous observations on LSCO \cite{Padilla-05}.

The  optical scattering rate $\tau^{-1}(\omega)$  in the right panels of figure \ref{S-ExtDrude}
exceeds the frequency $\omega$. As already reported for other cuprates \cite{Timusk-99,Basov-05}, this  violates the basic assumption of the Fermi-liquid theory $1/\tau(\omega)<\hbar\omega$,
which is demanded for the coherence of quasiparticles.
The overall magnitude of the scattering rate is remarkably enhanced
over the whole frequency range in the $p = 0.10$ material. The change of slope at $\sim1000$ cm$^{-1}$, particularly  evident at $p = 0.10$, which in the two component model identifies the MIR band, in the extended Drude approach  indicates the opening of a pseudogap \cite{Timusk-99} in the density of states, which  does not appear directly in the $\sigma_1 (\omega)$.

\subsection{The spectral weight} 

In the literature of metallic cuprates, the $ab$-plane $\sigma_1(\omega)$
has been modeled  either by the extended Drude approach or by the
multi-component picture described in the preceding sections.  However,  information on their peculiar electrodynamics and especially on the role played by the correlation effects can be also provided by a model-independent quantity,  the optical spectral weight

\begin{equation}
W(\Omega,T)=\int_{0}^{\Omega}\sigma_1(\omega,T)\mathrm{d}\omega
\label{SW1}
\end{equation}

For $\Omega\rightarrow\infty$, according to the ordinary sum rule,  $W = \pi ne^2/4m_b$ is independent of $T$. Therein, $n$, $e$, and $m_b$ are the carrier density, charge, and band mass, respectively. 
If instead $\Omega$ is set close to the minimum in the reflectivity, which provides a rough estimate of the screened plasma frequency $\Omega_p\simeq$ 8000 cm$^{-1}$, one 
obtains a sum rule \emph{restricted} to the majority of the carriers in the conduction band \cite{Benfatto-05}.
 
\begin{figure}[t]
\begin{center}
\leavevmode    
\epsfxsize=8.5cm \epsfbox {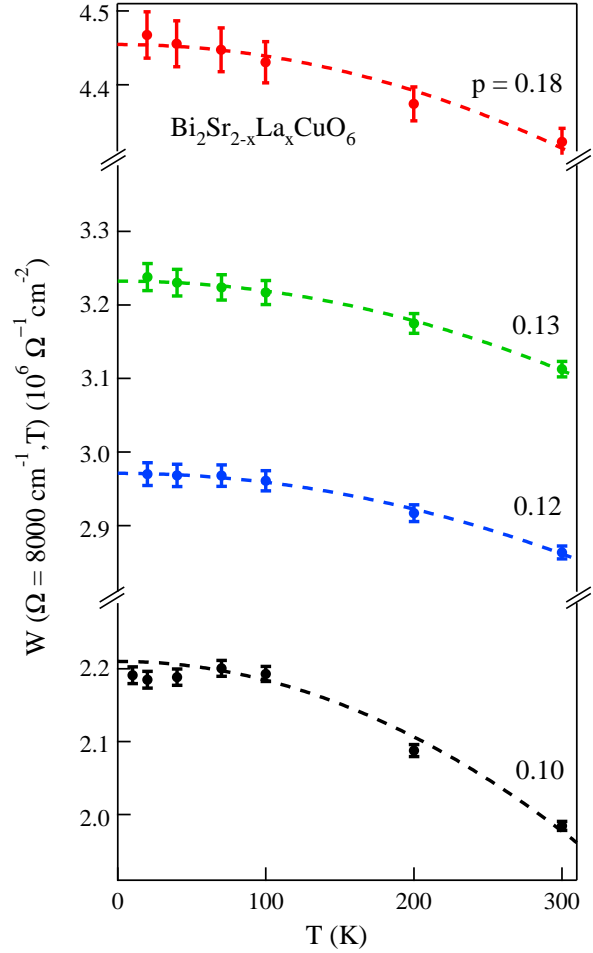}
\end{center}
\caption{Temperature dependence of the optical spectral weight $W(\omega_p)$ calculated at the screened plasma frequency $\omega_p$ in four Bi$_2$Sr$_{2-x}$La$_x$CuO$_6$ samples.
The dashed lines indicate the fits  to data of equation \ref{SW3}.}
\label{W-vs-T}
\end{figure}

The interest in the optical spectral weight is mostly due to its relation with the kinetic energy \cite{Hirsch-02}. Indeed \cite{Benfatto-05}:

\begin{equation}
W(\Omega_p,T)=\frac{\pi e^2}{\hbar^2V}\frac{1}{N}\sum_{\mathbf{k},s}\frac{\partial^2\varepsilon_{\mathbf{k}}}
{\partial k_i^2}n_{\mathbf{k},s} 
= -\frac{\pi e^2 a^2}{n_d\hbar^2 V}K(T)
\label{SW2}
\end{equation}

\noindent
where the sum is performed over $N$ momenta $\mathbf{k}$ of the first Brillouin zone, $\varepsilon_{\mathbf{k}}$ is the dispersion, $n_{\mathbf{k},s}$ the occupation number for a given $\mathbf{k}$ state with spin $s$, and $V$ the unit-cell volume.
The last relation in equation \ref{SW2}, where $K$ is the average kinetic energy and $n_d$ is the dimensionality of the system, only holds for a nearest-neighbor, tight-binding, single-band model.
Therefore, using the Sommerfeld expansion of the Fermi distribution function \cite{Boris-04} at the first order, one obtains for single-band tight-binding metals 

\begin{equation}
W(\Omega_p,T)\simeq W_0-B(\Omega)T^2
\label{SW3}
\end{equation}

\noindent
This equation is verified in gold \cite{Ortolani-05}, whose conduction band is indeed described by a one-band (the $6s$ band) tight-binding approximation \cite{Ashcroft-76}. As $W_0 \propto n/m_b$, it is related to the hopping rate $t_0$ by the general bandwidth relation $1/m_b\sim t_0$.
Furthermore, the "thermal" coefficient $B(\Omega)$ is crucially related to the density of states at the Fermi energy
$N(\epsilon_F)$, which depends on $n$ and on the details of the band structure,
but in general it will be inversely proportional to the bandwidth, and
hence to $t_0$. We thus have, for a conventional tight-binding metal like gold, $W_0\sim t_0$ and $B\propto N(\varepsilon_F)\sim t_0^{-1}$.

In cuprates, the $T^2$ behavior in equation \ref{SW3}  has been confirmed in BSCCO for $\Omega=\Omega_p$  \cite{Molegraaf-02,Santander-03} and in  LSCO for a wide frequency and doping range  \cite{Ortolani-05}. For the present  BSLCO crystals, the behavior of $W(\Omega_p,T)$ is displayed in figure \ref{W-vs-T}. It follows the $T^2$ dependence between $T=300$ K and the minimum $T>T_c$,
over the whole metallic side of its phase diagram, except
the underdoped material with $p = 0.10$ below 70 K, due probably to the
weak low-$T$ divergence of the in-plane resistivity reported in figure \ref{Rho}.
The error bars in figure \ref{W-vs-T} have been calculated
by assuming an experimental uncertainty of 1\% on the overall level of the measured $R(\omega)$.
Moreover, in a given sample, both the $T^2$ dependence of $W$ and
the value of  $B$ are the same within errors between $\Omega=\Omega_p/2$ and $\Omega=3\Omega_p/2$.

\begin{table} [t]
\begin{center}
\begin{tabular}{cccc}
\hline \hline 
Metal & $b(\Omega_p)$ & $\rho (300 K)$ & $\Omega_p$ \\ 
      & ($10^{-8}$ K$^{-2}$) & ($m\Omega$ cm) & (cm$^{-1}$) \\ 
\hline
{\it Correlated} & & & \\

Bi$_2$Sr$_{1.8}$La$_{0.2}$CuO$_6$ & 35  & 0.37 & 8000 \\
Bi$_2$Sr$_{1.6}$La$_{0.4}$CuO$_6$ & 40  & 0.50 & 8000 \\
Bi$_2$Sr$_{1.4}$La$_{0.6}$CuO$_6$ & 41  & 0.85 & 8000 \\
Bi$_2$Sr$_{1.3}$La$_{0.7}$CuO$_6$ & 41  & 1.2 & 8000 \\
Bi$_2$Sr$_{1.2}$La$_{0.8}$CuO$_6$ & 117 & 1.6 & 8000 \\
La$_{1.88}$Sr$_{0.12}$CuO$_4$ \cite{Ortolani-05} & 35 & 0.50 & 6800 \\ 
La$_{1.74}$Sr$_{0.26}$CuO$_4$ \cite{Ortolani-05} & 25 & 0.17 & 6800 \\ 
Bi$_2$Sr$_{2}$CaCu$_2$O$_8$ & 20 & 0.24 & 10000 \\
V$_2$O$_3$ \cite{Baldassarre-08, Baldassarre-09} & 160 & 0.37 & 8000 \\
\hline
{\it Uncorrelated} & & & \\

CaAlSi \cite{CaAlSi}	& 5.0 & 0.14 & 8000 \\ 
Au \cite{Ortolani-05} & 1.3 & 0.03 & 20500 \\ 
\hline \hline
\end{tabular}
\caption{Normalized coefficient $b(\Omega_p)=B(\Omega_p)/W_0$ of the $T^2$ term in the spectral weight
of the present samples (BSLCO and BSCCO) compared with corresponding data reported in the literature for different metallic systems.
Both the resitivity at 300 K and the screened plasma frequency of each material are reported for reference.}
\label{b-table}
\end{center}
\end{table}

\begin{figure}[t]
\begin{center}
\leavevmode    
\epsfxsize=8.5cm \epsfbox {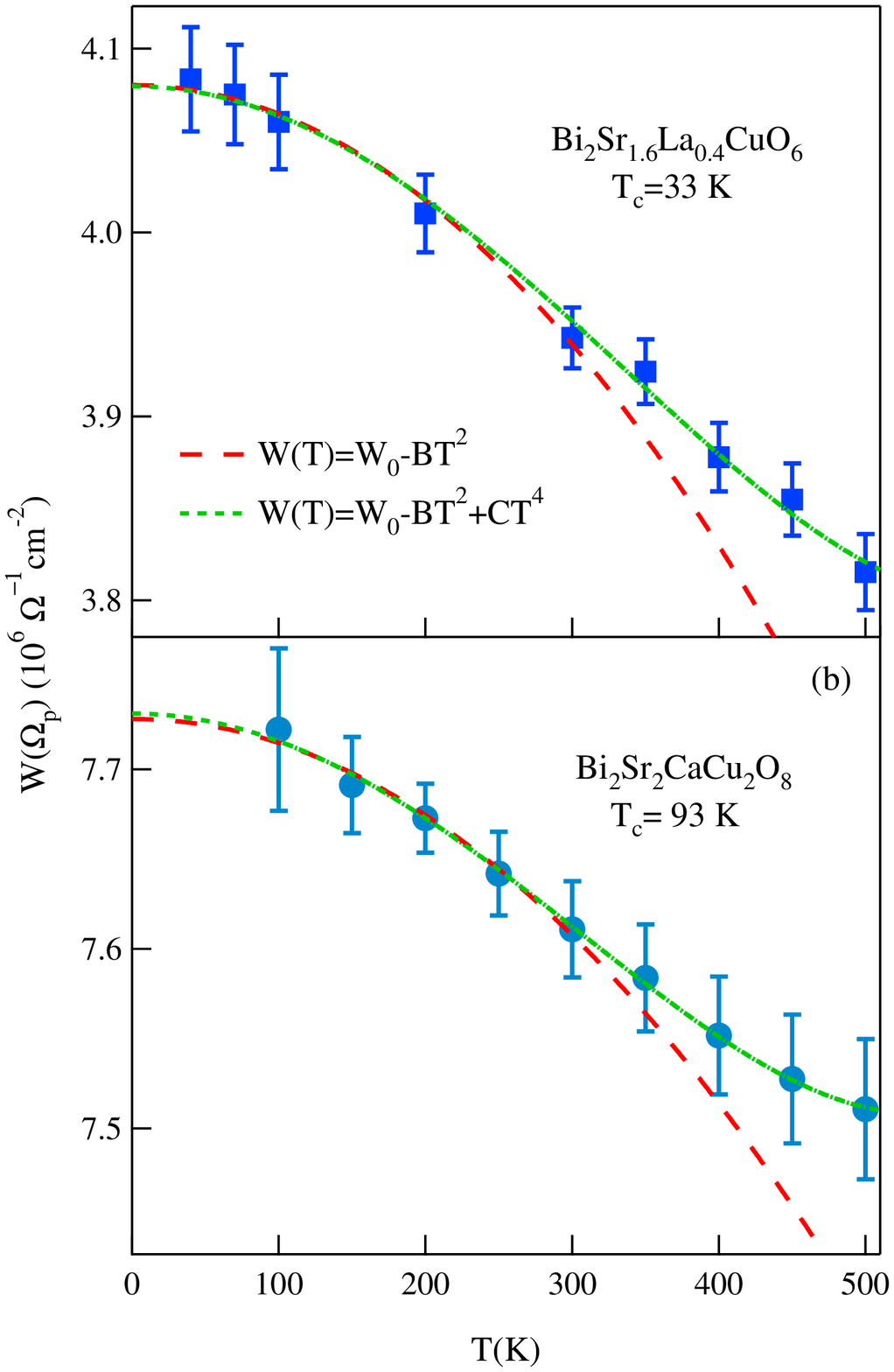}
\end{center}
\caption{Temperature dependence of the optical spectral weight $W(\Omega_p)$ for  Bi$_2$Sr$_{2-x}$La$_x$CuO$_6$ 
and  Bi$_2$Sr$_{2}$CaCu$_2$O$_8$, both at optimum doping. 
The dashed lines are the fits  to data of equation \ref{SW2}, dotted lines the fits  to data of equation \ref{SW4}.}
\label{W-highT}
\end{figure}

The frequency-dependent \emph{thermal response} $B(\Omega)$ can be evaluated at any
$p$ from fits like those in figure \ref{W-vs-T}. The resulting values are reported in table \ref{b-table}
in terms of normalized coefficients $b(\Omega_p)=B(\Omega_p)/W_0$. Therein, they are compared with those evaluated for other cuprates, for the Mott-Hubbard
model system V$_2$O$_3$ and for two uncorrelated metals.
As one can see, $b(\Omega_p)$ in BSLCO is comparable with that of the other cuprate families
(except in the strongly underdoped $x=0.8$ material),
while it is  larger  by roughly one order of magnitude than that of uncorrelated metals  .
The $b$ of V$_2$O$_3$ is even larger, in average, than that of cuprates.
Table 3 also indicates that high values of $b$, corresponding to a strong $T$-dependence of $W$, are poorly related to the resistivity $\rho$ of the metal and are not related at all to its plasma frequency.
These results are not
surprising in view of the strong correlation effects  in the latter compounds. Indeed, the motion of a hole can in principle result in a double electron occupancy on Cu sites, which costs the Hubbard energy $U$. The modification with $T$ of the filling of the
Hubbard bands  produces changes in $\sigma_1(\omega)$ up to an energy $U$ and this explains why $b(\Omega) \neq 0$ for $\Omega >> \Omega_p$. This relation has been quantitatively supported by Dynamical Mean Field (DMFT) calculations for LSCO, which predict the correct values of $b(\Omega_p$) only if one takes into account the effect of electronic correlations \cite{Toschi-05}.
The thermal response $B(\Omega_p)$, or $b(\Omega_p$), therefore  probes  experimentally
the correlation effects at finite $T$ and out of half-filling.

In order to verify both the accuracy of the Sommerfeld expansion and the effect of correlations at higher temperatures, we have extended the measurements on both the crystals at optimum doping (BSLCO with $x$ = 0.4 and   BSCCO) up to 500 K (see  figure \ref{sigma-metal}). The resulting spectral weight $W(\Omega_p)$, calculated at the plasma edge $\Omega_p$ ( $\simeq8000$ cm$^{-1}$ in BSLCO and $\simeq10000$ cm$^{-1}$ in BSCCO) is shown in figure \ref{W-highT}.

Therein, the $T^2$ dependence predicted by equation (\ref{SW3}) well fits the measured $W$  only for $T\lesssim300$ K, while a clear deviation appears above room temperature for both samples.
One may wonder whether this is due to the system approaching the I-R limit, where the quasiparticle picture breaks down due to the electron mean free path $\ell$ becoming comparable with the lattice constant $a$  \cite{Gunnarsson-03}. However, the calculated \cite{prl10}  $\sigma_{dc}^{I-R}$  (marked by arrows in figure \ref{sigma-metal}) shows that, even at 500 K, in both systems $\sigma_1(\omega)$ for $T \to 0$ is far from the Ioffe-Regel limit. 

On the contrary, the deviation of $W(\Omega_p, T)$ from the $T^2$ behavior is satisfactorily reproduced at all temperatures if one includes the $T^4$ term of the Sommerfeld expansion  (dotted lines in figure \ref{W-highT}), namely if $W$ is fitted to 

\begin{equation}
W(\Omega_p,T)=W_0-BT^2+CT^4
\label{SW4}
\end{equation}

\noindent
We obtained $b(\Omega_p)=B/W_0\simeq4.0\cdot10^{-7}$ K$^{-2}$ in BSLCO and $b(\Omega_p)\simeq2.0\cdot10^{-7}$ K$^{-2}$ in BSCCO. Therefore, $c(\Omega_p) = 6.1\cdot10^{-13}$ ($3.6\cdot10^{-13}$) K$^{-4}$ and $c(\Omega_p)/b(\Omega_p) = (1.5 \pm 0.4)\cdot10^{-6}$ (($1.8\pm 0.5)\cdot10^{-6}$) K$^{-2}$ for BSLCO (BSCCO).
In order to check the generality of such behavior,  the cutoff frequency $\Omega$
in equation \ref{SW1} was varied from $\Omega_p/2$ to $3\Omega_p/2$, and  deviations from the $T^2$ dependence quite similar to those in figure \ref{W-highT} were always found. 
DMFT calculations \cite{prl10} show that either the absolute values of the above coefficients and the $T^4$ behavior are influenced by strong correlations, which then produce major effects also at high temperature. 


\section{The superconducting phase} 
\subsection{The optical conductivity}

\begin{figure}[t]
\begin{center}
\leavevmode    
\epsfxsize=8.5cm \epsfbox {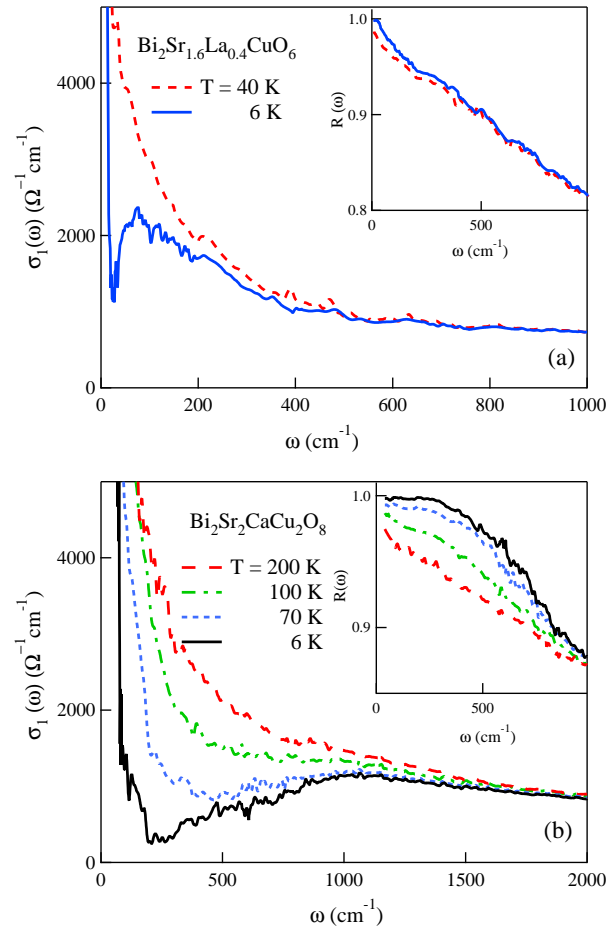}
\end{center}
\caption{The $ab$-plane optical conductivity and reflectivity (inset)  of optimally-doped
Bi$_2$Sr$_{1.6}$La$_{0.4}$CuO$_6$ (a) and Bi$_2$Sr$_2$CaCu$_2$O$_8$ (b) above and below their $T_c$.}
\label{sigma-sup}
\end{figure}

We have also studied (down to 10  cm$^{-1}$), both optimally-doped samples (Bi$_2$Sr$_{1.6}$La$_{0.4}$CuO$_6$ and BSCCO) below their  $T_c $ (33 and 93 K, respectively). The data of BSCCO are similar to those reported previously for the same material \cite{Hwang-07, Carbone-06}. 
The low-frequency $\sigma_1(\omega)$ and  $R(\omega)$  of both compounds are displayed in  figure \ref{sigma-sup},   above and below $T_c$. In both panels, below $T_c$, $\sigma_1(\omega)$ exhibits the partial opening of a gap and residual Drude conductivity below 20 cm$^{-1}$ in BSLCO and 100 cm$^{-1}$ in BSCCO. 
A similar behavior  is usually observed in $d$-wave superconductors and can be attributed to the infrared average of  the anisotropic density of states in the momentum space. The residual Drude may then come from the ungapped zones  of the Fermi surface.

\begin{figure}[t]
\begin{center}
\leavevmode    
\epsfxsize=12cm \epsfbox {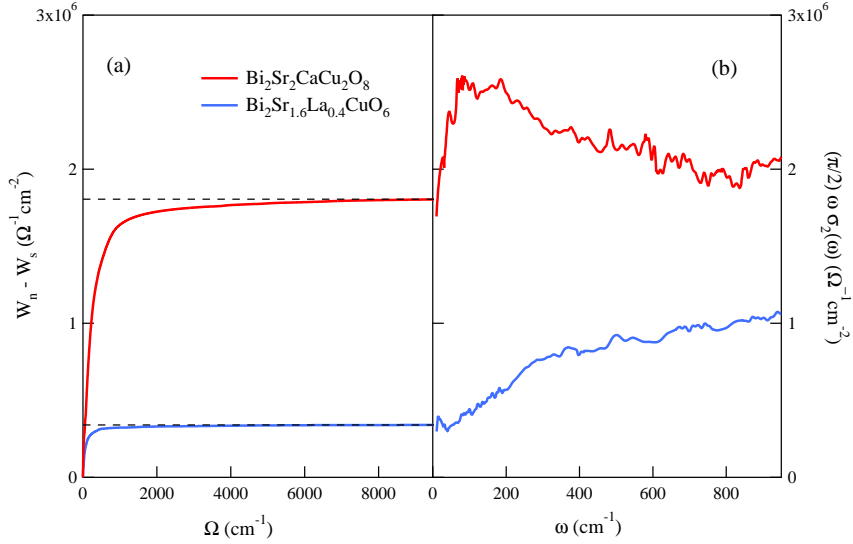}
\end{center}
\caption{(a) The difference $W_n-W_s$ between the spectral weight calculated from
$\sigma_1(\omega)$ at $T\gtrsim T_c$ and that at 6 K (see text) is plotted vs. the integration limit
$\Omega$ for optimally-doped Bi$_2$Sr$_2$CaCu$_2$O$_8$ and Bi$_2$Sr$_{1.6}$La$_{0.4}$CuO$_6$.
(b) $(\pi/2)\omega\sigma_2(\omega)$ in the far-IR for both samples at 6 K.}
\label{SW-SC}
\end{figure}

Basing on the Ferrel-Glover-Tinkham  sum rule (FGT) one usually obtains the  spectral weight of the $\delta$ function which condenses at $\omega = 0$, namely the superfluid density, through

\begin{equation}
\frac{\omega^2_{pS}}{8} = W_n-W_s=\int_{0}^{\Omega}\left[\sigma_{1n}(\omega)-\sigma_{1s}(\omega)\right]\mathrm{d}\omega
\label{F-G-T}
\end{equation}

\noindent
with $\Omega\gtrsim6\Delta$.
$W_n$ is calculated at the lowest $T>T_c$ (\emph{i.e.}, 40 K) and $W_s$ is obtained at $T=6$ K,  well below $T_c$ in both materials. The result is plotted vs. $\omega$ in figure \ref {SW-SC}(a).
An alternative estimate of $\omega^2_{pS}$ is given by the imaginary part $\sigma_2(\omega)$ of
the complex conductivity  through the London equation for the electrodynamics of superconductors, which gives

\begin{equation}
\lim_{\omega\rightarrow0}4\pi\omega\sigma_2=\omega_{pS}^2
\label{London-Limit}
\end{equation}

\begin{table} [t]
\begin{center}
\begin{tabular}{|c|c c c|}
\hline
& Bi-2212 & BSLCO & LSCO \\
\hline \hline
$T_c$ (K) & 93 & 33 & 41 \\
$\lambda_L^{IR}$ (nm) & 137 & 291 & 215 \\
$\lambda_L^{\mu SR}$ (nm) & 169 & 297 & 246 \\
\hline
\end{tabular}
\caption{The London penetration depth $\lambda_L^{IR}$ measured by infrared spectroscopy in  optimally-doped BSCCO, BSLCO (present work) and LSCO \cite{Ortolani-05}, is compared with the corresponding $\lambda_L^{\mu SR}$ measured by muon spin relaxation \cite{Russo-07,Kadono-04}.}
\label{Table-PenDepth}
\end{center}
\end{table}

\noindent
In figure \ref{SW-SC}-b this limit, divided by 8, coincides \cite{Molegraaf-02,Santander-03} with the difference $W_n-W_s$ provided that $\sigma_1(\omega)$ is integrated in
Eq. \ref{F-G-T} up to $\Omega\gtrsim 2000$ cm$^{-1}$ in BSLCO and $\Omega\gtrsim 8000$ cm$^{-1}$ in BSCCO.  Therefore, as already reported for other
hole-doped superconducting cuprates \cite{Ortolani-05, Molegraaf-02},
the energy range involved in the FGT sum rule is much larger than the Bardeen-Cooper-Schrieffer (BCS) prediction  $\Omega \simeq 6\Delta$ (4$\Delta$) where $\Delta$ is the superconducting gap  in the dirty (clean) limit. In optimally doped BSLCO, $6\Delta\sim300$ cm$^{-1}$, while in BSCCO $6\Delta\sim1500$ cm$^{-1}$ \cite{Yurgens-03}.
One may notice that $(\pi/2)\omega\sigma_2$ in  figure \ref {SW-SC} should be fairly constant for any $\omega$ in the region of the superconducting gap. Its frequency dependence in the far IR is due the presence of a residual Drude-like absorption in figure \ref{sigma-sup}.

From $W_n-W_s$ one can also extract the London penetration depth 

\begin{equation}
\lambda_L^{IR}=\sqrt{\frac{mc^2}{4\pi n_se^2}}
\label{depth}
\end{equation}

\noindent
The values thus obtained for BSLCO and BSCCO are reported in Table \ref{Table-PenDepth},
where $\lambda_L^{IR}$ of optimally-doped LSCO from Ref. \cite{Ortolani-05} is also shown for comparison.
As one can see, they are found in good agreement with recent \cite{Kadono-04, Russo-07} measurements of
$\lambda_L^{\mu SR}$, extrapolated to $T=0$, by muon spin spectroscopy.
Table \ref{Table-PenDepth} also shows the strong increase in the London penetration depth as  $T_c$ decreases from 93 K to 33 K.


\section{Concluding remarks} 

In the present work we have measured the $ab$-plane reflectivity of  single crystals belonging to three different one-layer cuprate families. This allowed us to analyze the  infrared response  of the doped Cu-O plane throughout its  $p,T$ phase diagram, from $p\simeq$ 0 to $p$ = 0.18, and for $T$ varying from 6 to 500 K.  In the insulating phase we have measured the low-temperature phonon spectrum  of SCOC, YCBCO, and BSLCO, which has been compared with previous observations at room temperature. At extreme hole dilution ($p \simeq$ 0 and $p$ = 0.005), a narrow peak at 1570 cm$^{-1}$  (195 meV) was  observed  for the first time in a cuprate, which increases in intensity with doping. It was assigned to the  ground state of a hole bound to a Sr defect. By using that energy value and a simple Mott-transition model, we could predict a critical value for $p_{IMT}$ at the insulator-to-metal transition which differs by a factor of 2 from that really observed in another compound, BSLCO.  Considering that in simpler semiconductors the disagreement between the Mott criterion and the observations is generally much larger, we consider this result as a confirmation that the IMT in cuprates is driven by a conventional transformation of isolated-impurity levels into a conduction band at a critical $p_{IMT}$. This is indeed what we have observed by monitoring the optical conductivity at low temperature for increasing doping. The level at 195 meV broadens into a FIR band whose low-energy edge forms an insulating gap which closes as either $p$ or $T$ increase. In the latter case however,  in the far infrared instead of the true Drude peak a flat band appears, which is suggestive of thermally-activated incoherent hopping. 
The IMT occurs at low $T$ only when  the FIR band turns into a Drude term, which continues to develop when doping further increases. During the whole process, a mid-infrared band well distinguished from the Drude absorption also softens,  from 3000 cm$^{-1}$ at $p \simeq$ 0 to about 1000  cm$^{-1}$ at $p_{IMT}$. However, it does not seem to  participate actively to the IMT, and the fact that it originates in the AF phase may imply that it is magnetic in nature, as proposed for similar infrared bands by recent theoretical models.

In the metallic phase, an analysis of the optical conductivity alternative to the MIR+Drude model has been performed in terms of the extended-Drude approach.  The presence of a pseudogap may be  assumed for the strongly underdoped compound only, where however the clear separation between the Drude and MIR contributions makes the model unreliable.
We have then analyzed in detail a model-independent quantity, the infrared spectral weight $W(T)$. Below 300 K this depends on temperature like $W_0 - BT^2$, as predicted by the Sommerfeld expansion of the carrier kinetic energy truncated at the first order, and as already observed in LSCO and other cuprates. Like in LSCO, in BSLCO there are however two independent scales, which differ in energy by approximately one order of magnitude, that of $W_0$ and that of the $T^2$-coefficient $B$. Moreover, this remains nearly constant with frequency well above the shielded plasma frequency. DMFT calculations showed that these effects are related to the strong correlations present in the Cu-O plane. An interesting effect comes out when extending the study of $W(T)$ to temperatures higher than 300 K. In both the superconductors examined, the single-layer BSLCO and the bilayer BSCCO, $W(T)$ is still well described by the Sommerfeld expansion,  provided however that  the $T^4$ term is included. 

In the superconducting phase BSLCO behaves similarly to BSCCO, once the gap energy, the width of the residual Drude peak below $T_c$, and the superconducting carrier density are scaled down in consideration of its lower critical temperature. As previously reported for other cuprates, also in BSLCO the FGT sum rule is fulfilled only if the integration of the conductivity is extended to frequencies much beyond the BCS limit of 6$\Delta$.


\section*{Acknowledgments}
We are indebted to A. Erb for providing the YCBCO sample and to L. Baldassarre for measuring its infrared spectra. 
We acknowledge the Helmholtz-Zentrum Berlin -  Electron storage ring BESSY 
II for provision of synchrotron radiation at beamline IRIS. The research leading to these results has received funding from the 
European Community's Seventh Framework Programme (FP7/2007-2013) under grant 
agreement n.226716.


\end{document}